\documentclass[secthm]{elsart}

\usepackage{graphicx}
\usepackage{amsmath, amssymb}
\usepackage{mathrsfs}
\usepackage{color}
\usepackage{colortbl}

\usepackage{ifpdf}
\ifpdf
\usepackage[%
  pdftitle={Instructions for use of the document class
    elsart},%
  pdfauthor={Simon Pepping},%
  pdfsubject={The preprint document class elsart},%
  pdfkeywords={instructions for use, elsart, document class},%
  pdfstartview=FitH,%
  bookmarks=true,%
  bookmarksopen=true,%
  breaklinks=true,%
  colorlinks=true,%
  linkcolor=blue,anchorcolor=blue,%
  citecolor=blue,filecolor=blue,%
  menucolor=blue,pagecolor=blue,%
  urlcolor=blue]{hyperref}
\else
\usepackage[%
  breaklinks=true,%
  colorlinks=true,%
  linkcolor=blue,anchorcolor=blue,%
  citecolor=blue,filecolor=blue,%
  menucolor=blue,pagecolor=blue,%
  urlcolor=blue]{hyperref}
\fi

\newcommand{\set}[1]{\left\{#1\right\}}
\newcommand{\Set}[2]{\set{#1\ \vert\ #2}}

\newtheorem{obs}[thm]{Observation}
\newtheorem{algo}[thm]{Algorithm}
\definecolor{darkmagenta}{rgb}{0.5,0,0.5}

\begin{document}

\pagestyle{headings}

\begin{frontmatter}

\title{Characterizations of probe interval graphs}

\author{Shamik Ghosh\corauthref{cor}}
\corauth[cor]{Corresponding author.}
\ead{sghosh@math.jdvu.ac.in},
\author{Maitry Podder}
\ead{maitry\_podder@yahoo.co.in}

\address{Department of Mathematics, Jadavpur University, Kolkata - 700 032, India.}

\author{Malay K. Sen}
\ead{senmalay@hotmail.com}

\address{Department of Mathematics, North Bengal University,
   Darjeeling, West Bengal, India, Pin - 734 430.}


\begin{abstract}
In this paper we obtain several characterizations of the adjacency matrix of a probe interval graph. In course of this study we describe an easy method of obtaining interval representation of an interval bipartite graph from its adjacency matrix. Finally, we note that if we add a loop at every probe vertex of a probe interval graph, then the Ferrers dimension of the corresponding symmetric bipartite graph is at most $3$. 
\end{abstract}


\end{frontmatter}

\section{Introduction}

An undirected graph $G=(V,E)$ is an {\em{interval graph}} if it is the intersection graph of a family of intervals on the real line in which each vertex is assigned an interval and two vertices are adjacent if and only if their corresponding intervals intersect. The study of interval graphs was spearheaded by Benzer \cite{B} in course of his studies of the topology of the fine structure of genes. Since then interval graphs and their various generalizations were studied thoroughly. Also advances in the field of molecular biology, and genetics in particular, solicited the need for a new model. In \cite{Z}, Zhang introduced another generalization of interval graphs called probe interval graphs, in an attempt to aid a problem called cosmid contig mapping, a particular component of the physical mapping of DNA. A {\em{probe interval graph}} is an undirected graph $G=(V,E)$ in which the set of vertices $V$ can be partitioned into two subsets $P$ and $N$ (called probes and nonprobes respectively) and there is an interval (on the real line) corresponding to each vertex such that vertices are adjacent if and only if their corresponding intervals intersect and at least one of the vertices belongs to $P$. Now several research works are continuing on this topic and some special classes of it \cite{BL,CK,GL,JS,MWZ,LS}. In fact, Golumbic and Trenk have devoted an entire chapter on probe interval graphs in their recent book \cite{GT}. Moreover, motivated by the definition of probe interval graphs, genrally, the concept of probe graph classes has been introduced. Given a class of graphs $\mathscr{G}$, a graph $G$ is a {\em probe graph of} $\mathscr{G}$ if its vertices can be partitioned into a set $P$ of probes and an independent set $N$ of nonprobes such that $G$ can be extended to a graph of $\mathscr{G}$ by adding edges between certain nonprobes. In this way, many more probe graph classes have been defined and widely investigated, eg., probe split graphs, probe chordal graphs, probe tolerance graphs, probe threshold graphs and others \cite{BGL,CKKLP,RB,RLB}.

Among all such studies nothing has been said about the nature of adjacency matrices of probe interval graphs until now. In this paper we will present three characterizations of the adjacency matrix of a probe interval graph. The first one is in this section and two others are in section 3. In section 2, we describe an easy method of obtaining interval representation of an interval bipartite graph from its adjacency matrix. Moreover, we prove that if we add a loop at every probe vertex of a probe interval graph, then the Ferrers dimension of the corresponding symmetric bipartite graph is at most $3$.

We first note that any interval graph $G=(V,E)$ is a probe interval graph with probes $P$ and nonprobes $N$, where $N$ is any independent set (possibly singleton) of $G$ and $P=V\smallsetminus N$. Certainly the converse is false, as $C_4$ is a probe interval graph which is not an interval graph. So probe interval graphs generalize the class of interval graphs. Further generalizations lead to the following concepts. An undirected graph $G=(V,E)$ is an {\em{interval split graph}} \cite{Br} if the set of vertices $V$ are partitioned into two subsets $U_1$ and $U_2$ such that the subgraph induced by $U_1$ is an interval graph and $U_2$ is an independent set. Every probe interval graph is an interval split graph, as $N$ is an independent set and the subgraph induced by $P$ is an interval graph. Again interval bipartite graphs (cf.~\S 2) are generalized to interval $k$-graphs. An undirected graph with a proper coloring by $k$ colors is an {\em{interval $k$-graph}} \cite{Br} if each vertex corresponds to an interval on the real line so that two vertices are adjacent if and only if their corresponding intervals intersect and they are of different colors. Brown \cite{Br} showed that any $k$-chromatic probe interval graph is an interval $k$-graph. Also, since interval $k$-graphs are weakly chordal\footnote{An undirected graph $G$ is {\em weakly chordal} if neither $G$ nor its complement $\bar{G}$ contains an induced cycle of length greater than $4$.} and hence perfect,\footnote{An undirected graph $G$ is {\em perfect} if for every induced subgraph $H$ of $G$, the chromatic number of $H$ is equal to its maximal clique size.} we have that probe interval graphs are also weakly chordal and perfect. While comparing two graphs discussed earlier, Brown \cite{Br} made the comment:
``there are interval split graphs which are not interval $k$-graphs (e.g., $C_5$ or any cycle of length greater than or equal to $5$). The converse is not known -- but has not received much attention.'' The following example shows that there are interval $k$-graphs which are not interval split graphs.

\begin{exmp} {\em Consider the graph $G=K_{2,2,2}$, which is an interval $3$-graph. But it is not an interval split graph, since it has only $3$ independent sets, namely, $\set{a,d},\set{b,c}$ and $\set{x,y}$. For each such choice, the other $4$ vertices induce the subgraph $C_4$ which is not an interval graph.}

\begin{figure}[h]
\begin{center}
\includegraphics*[scale=0.4]{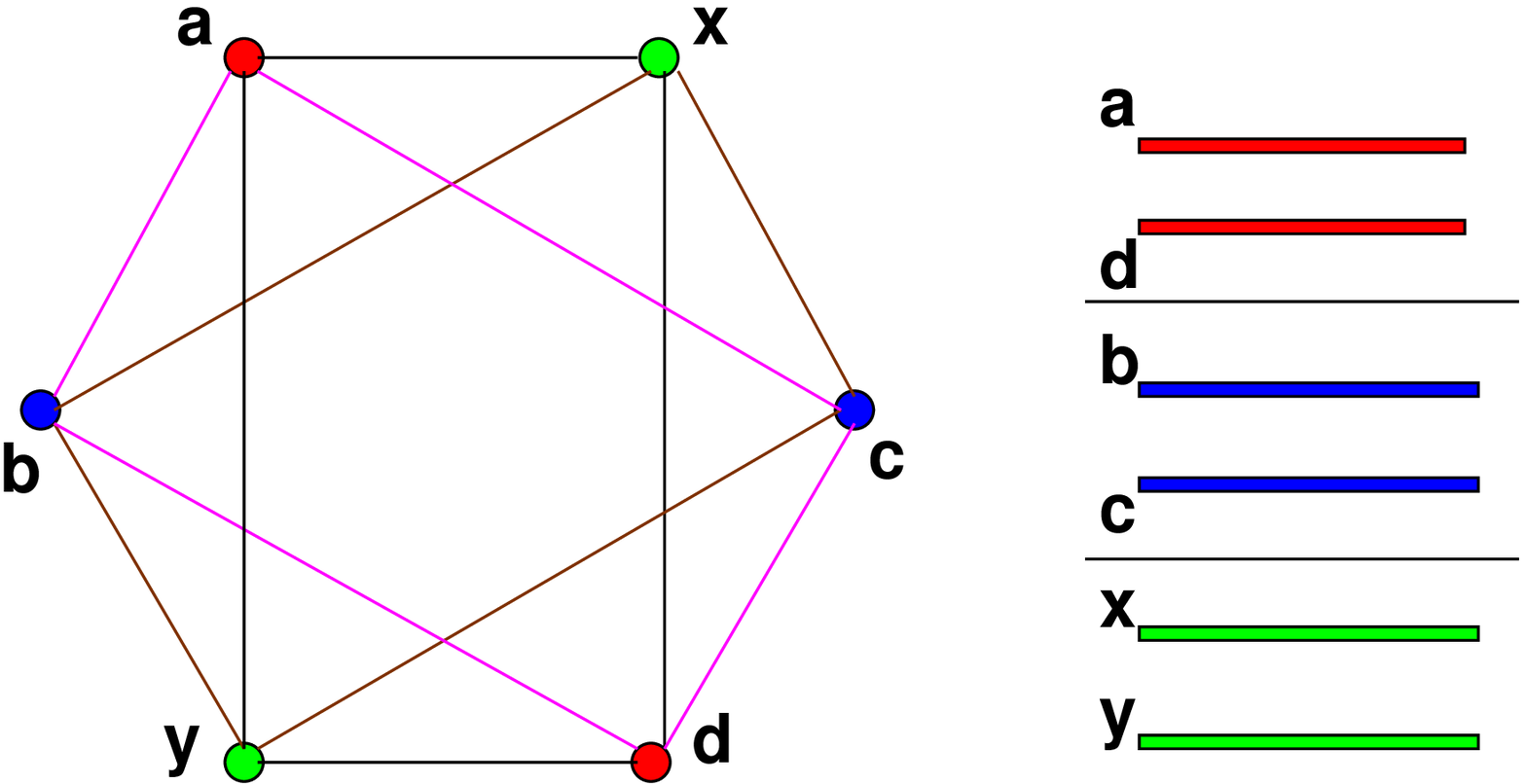}
\end{center}
\end{figure}
\end{exmp}

Now the class of probe interval graphs lies in the intersection of the class of interval split graphs and the class of interval $k$-graphs but there are examples (Brown presented one such in \cite{Br}) which are both interval split graphs and interval $k$-graphs but not probe interval graphs. Thus the following is an interesting open problem to study.

\begin{prob}
Characterize the class of graphs which are both interval split graphs and interval $k$-graphs.
\end{prob}

Regarding forbidden subgraph characterizations, Brown \cite{Br} showed that interval $k$-graphs and hence probe interval graphs are ATE-free\footnote{An {\em asteroidal triple of edges} (ATE) in an undirected graph $G$ is a set of three edges such that for any two there exists a path in $G$ that avoids the neighborhood of the third edge.}, while Sheng \cite{LS} characterized cycle free probe interval graphs in terms of six forbidden subgraphs (trees). Among the other characterizations, Brown \cite{Br} and Zhang \cite{Z} generalized the well known \cite{GH,G} result that an undirected graph is an interval graph if and only if its maximal cliques are consecutively ordered\footnote{A set of distinct induced subgraphs $\mathcal{G}=\set{G_1,G_2,\ldots,G_t}$ of a graph $G=(V,E)$ is {\em consecutively ordered} when for each $v\in V$, if $i<j<l$ and $v\in G_i\cap G_l$, then $v\in G_j$.}. Brown \cite{Br} proved that if $G=(V,E)$ is an undirected graph with an independent set $N\subseteq V$, then $G$ is a probe interval graph with probes $P=V\smallsetminus N$ and nonprobes $N$ if and only if $G$ has an edge cover of quasi-cliques\footnote{A {\em quasi-clique} $Q$ in a probe interval graph $G=(P,N,E)$ is a set of vertices with all vertices of $Q\cap P$ are adjacent to each other and any vertex of $Q\cap N$ is adjacent to all vertices of $Q\cap P$.} that can be consecutively ordered. 

In the following we shall present the first characterization of adjacency matrices of probe interval graphs (cf. Observation \ref{o:char1}), which is simple and immediate. Let $G=(V,E)$ be a simple undirected graph. If we replace\footnote{This replacement is equivalent to add a loop at each vertex of $G$.} all principal diagonal elements\footnote{which are $0$ in the adjacency matrix of $G$.} of the adjacency matrix of $G$ by $1$, then this matrix is known as the {\em \label{'augmented'} augmented adjacency matrix} of $G$. Let $M$ be a symmetric $(0,1)$ matrix with $1$'s in the principal diagonal. Then $M$ is said to satisfy the {\em quasi-linear ones property} if $1$'s are consecutive right to and below the principal diagonal. It is known \cite{MR} that $G$ is an interval graph if and only if rows and columns of the augmented adjacency matrix of $G$ can be suitably permuted (using the same permutation for rows and columns) so that it satisfies the quasi-linear ones property.

\begin{defn} 
{\em Let $M$ be a symmetric $(0,1)$-matrix with $1$'s in the principal diagonal. Suppose $M$ contains a principal submatrix\footnote{We call a square submatrix $N$ of $M$ {\em{principal}} if the principal diagonal elements of $N$ are also principal diagonal elements of $M$.} $N$ which is an identity matrix. Denote all the zeros of $N$ by $X$. Then $M$ is said to satisfy the {\em{quasi-x-linear ones property}} if every $0$ right to the principal diagonal has only $0$ and $X$ to its right and every $0$ below the principal diagonal has only $0$ and $X$ below it.}  
\end{defn}

\vspace{1em} Now from the definition of a probe interval graph $G$ it follows that the graph obtained by adding edges to $G$ between the pairs of nonprobes whose intervals intersect is an interval graph with the same assignment of intervals to the vertices as in $G$. Conversely, let $G=(V,E)$ be an interval graph and $N\subseteq V$. Then the graph obtained by removing all the edges between any two vertices belonging to $N$ from $G$ is a probe interval graph with probes $P=V\smallsetminus N$ and nonprobes $N$. Thus for an undirected graph $G=(V,E)$ with an independent set $N$, if adding edges between some vertices of $N$ make it an interval graph, then the graph $G$ must be a probe interval graph with probes $P=V\smallsetminus N$ and nonprobes $N$. This simple observation leads to the following characterization of probe interval graphs:

\vspace{2em}\begin{obs}\label{o:char1}
Let $G=(V,E)$ be an undirected graph with an independent set $N\subseteq V$. Let $A(G)$ be the augmented adjacency matrix of $G$. Then $G$ is a probe interval graph with probes $P=V\smallsetminus N$ and nonprobes $N$ if and only if rows and columns of $A(G)$ can be suitably permuted (using the same permutation for rows and columns) in such a way that it satisfies the quasi-x-linear ones property. 
\end{obs}

\vspace{-1.5em} \section{Interval representations of interval bipartite graphs}

\vspace{-1.25em} An {\em{interval bipartite graph}} (in short, {\em interval bigraph}) is a bipartite graph $B=(X,Y,E)$ with bipartation $(X,Y)$, representable by assigning each vertex $v\in X\cup Y$ an interval $I_v$ (on the real line) so that two vertices $x\in X$ and $y\in Y$ are adjacent if and only if $I_x\cap I_y\neq\emptyset$ \cite{HKM}. Since $X$ and $Y$ are independent sets in $B$, here we only consider the submatrix of the adjacency matrix of $B$ consisting of the rows corresponding to one partite set and the columns corresponding to the other. This submatrix is known as the {\em{biadjacency matrix}} of $B$. A bipartite graph $B$ is an interval bigraph if and only if the rows and columns of the biadjacency matrix of $B$ can be (independently) permuted so that each $0$ can be replaced by $R$ or $C$ in such a way that every $R$ has only $R$'s to its right and every $C$ has only $C$'s below it. Such a partition of zeros in the biadjacency matrix of $B$ is called an {\em{R-C partition}} of it \cite{SDRW}. Again a $(0,1)$-matrix $A$ has the {\em generalized linear ones property} if it has a stair partition\footnote{A {\em stair partition} of a matrix is a partition of its positions into two sets $(L,U)$ by a polygonal path from the upper left to the lower right, such that the set $L$ [$U$] is closed under leftward or downward [respectively, rightward or upward] movement \cite{SDW}.} $(L,U)$ such that the $1$'s in $U$ are consecutive and appear leftmost in each row, and the $1$'s in $L$ are consecutive and appear topmost in each column. For the biadjacency matrix $A$ of a bipartite graph $B$ this property is equivalent to having an R-C partition, i.e., $B$ is an interval bigraph if and only if the rows and columns of $A$ can be (independently) permuted so that the resulting matrix has the generalized linear ones property \cite{SDW}. Now there are many methods \cite{Mu,SDRW,W} of obtaining interval representation of an interval bigraph when the R-C partition of its biadjacency matrix is given. We present here another one for further use. 

\begin{defn}
{\em Let $B$ be an interval bigraph with the biadjacency matrix $A$, which is in R-C partition form. We insert some rows and columns in $A$, each of which has all the entries $X$ except the principal diagonal element which is $1$ such that $A$ is enhanced to a square matrix in which each $R$ is right to the principal diagonal and each $C$ is below the principal diagonal. Now replace each $X$ right to $R$ by $R$ and each $X$ below $C$ by $C$ and the rest by $1$. This matrix, say, $\tilde{A}$ is called a {\em diagonalized} form of $A$ and the above process of obtaining $\tilde{A}$ from $A$ will be called a {\em{diagonalization}} of $A$. We denote the bigraph whose biadjacency matrix is $\tilde{A}$ by $\tilde{B}$}\footnote{Note that $\tilde{B}$ is also an interval bigraph, as $\tilde{A}$ is still in R-C partition form and $B$ is an induced subgraph of $\tilde{B}$.}.
\end{defn}

An easy method of diagonalization is as follows. In the stair partition of $A$, if a step, parallel to rows [columns], is lying through $k$ columns [respectively, rows], then insert $k$ rows [respectively, columns] (as described previously) just above [respectively, after] the step. Accordingly we get a diagonalized matrix $\tilde{A}$ whose number of rows (as well as columns) is the sum of number of rows and columns of $A$. For practical purpose the number of insertions of rows and columns can be reduced as the following example shows.

\begin{exmp}\label{exmp:diag}
{\em Consider the following biadjacency matrix $A$ of an interval bigraph:}

\vspace{-1.5em}
$$\begin{array}{c|ccccc}
\textrm{\small Vertices} & x_1 & x_2 & x_3 & x_4 & x_5\\
\hline
y_1 & 1 & 1 & 1 & 0 & 0 \\
y_2 & 1 & 0 & 0 & 1 & 0 \\
y_3 & 0 & 0 & 0 & 1 & 0
\end{array}\hspace{1in} \begin{array}{c|ccccc}
\textrm{\small Vertices} & x_1 & x_2 & x_3 & x_4 & x_5\\
\hline
y_1 & 1 & 1 & 1 & R & \multicolumn{1}{c@{\,\vline}}{R}\\
\cline{2-4}
y_2 & 1 & C & \multicolumn{1}{c@{\,\vline}}{C} & 1 & \multicolumn{1}{c@{\,\vline}}{R}\\
\cline{5-5}
y_3 & C & C & C & \multicolumn{1}{c@{\,\vline}}{1} & \multicolumn{1}{c@{\,\vline}}{R}\\
\cline{2-6}
\end{array}$$
{\em A diagonalization of $A$ is given by} 
$$\begin{array}{c|cccccc}
\textrm{\small v} & x_1 & x_2 & x_3 & x_4 & x_6 & x_5\\
\hline
y_1 & 1 & 1 & 1 & R & X & R \\
y_4 & X & 1 & X & X & X & X \\
y_5 & X & X & 1 & X & X & X \\
y_2 & 1 & C & C & 1 & X & R \\
y_3 & C & C & C & 1 & 1 & R \\
y_6 & X & X & X & X & X & 1
\end{array}\hspace{1in}
\begin{array}{c|cccccc}
\textrm{\small v} & x_1 & x_2 & x_3 & x_4 & x_6 & x_5\\
\hline
y_1 & \textcolor{darkmagenta}{1} & 1 & 1 & R & R & R \\
y_4 & 1 & \textcolor{darkmagenta}{1} & 1 & 1 & 1 & 1 \\
y_5 & 1 & 1 & \textcolor{darkmagenta}{1} & 1 & 1 & 1 \\
y_2 & 1 & C & C & \textcolor{darkmagenta}{1} & 1 & R \\
y_3 & C & C & C & 1 & \textcolor{darkmagenta}{1} & R \\
y_6 & C & C & C & 1 & 1 & \textcolor{darkmagenta}{1}
\end{array}$$
\end{exmp}

\vspace{-1em}Now we present an algorithm to obtain an interval representation of an interval bigraph $B$.

\begin{algo} \label{alg:big1}
{\em {\small\bf Input:}\ Diagonalized matrix $\tilde{A}$ (of order $n\times n$ (say)), where $A$ is the biadjacency matrix (in R-C partition form) of an interval bigraph $B$.\\
{\small\bf Step I:}\ For each $i=1\textrm{ to }n$, define $a_i=i$ and $b_i=r$, where in the $i^{\textrm{th}}$ row the last $1$ appears in the $r^{\textrm{th}}$ column on or after the principal diagonal of $\tilde{A}$.\\
{\small\bf Step II:}\ For each $j=1\textrm{ to }n$, define $c_j=j$ and $d_j=s$, where in the $j^{\textrm{th}}$ column the last $1$ appears in the $s^{\textrm{th}}$ row on or after the principal diagonal  \mbox{of $\tilde{A}$}.\\
{\small\bf Output:}\ The closed intervals $[a_i,b_i]$ and $[c_j,d_j]$, which are corresponding to the $i^{\textrm{th}}$ row and $j^{\textrm{th}}$ column of $\tilde{A}$ respectively.}
\end{algo}

Using the above algorithm in the case of the interval bigraph considered in Example \ref{exmp:diag}, we have
$$\begin{array}{c|cccccc|c}
\textrm{\small Vertices} & x_1 & x_2 & x_3 & x_4 & x_6 & x_5 & \textrm{\small Intervals}\\
\hline
y_1 & \textcolor{darkmagenta}{1} & 1 & 1 & R & R & R & [1,3]\\
y_4 & 1 & \textcolor{darkmagenta}{1} & 1 & 1 & 1 & 1 & [2,6]\\
y_5 & 1 & 1 & \textcolor{darkmagenta}{1} & 1 & 1 & 1 & [3,6]\\
y_2 & 1 & C & C & \textcolor{darkmagenta}{1} & 1 & R & [4,5]\\
y_3 & C & C & C & 1 & \textcolor{darkmagenta}{1} & R & [5,5]\\
y_6 & C & C & C & 1 & 1 & \textcolor{darkmagenta}{1} & [6,6]\\
\hline
\textrm{\small Intervals} & [1,4] & [2,3] & [3,3] & [4,6] & [5,6] & [6,6] &
\end{array}$$
Finally removing newly inserted rows and columns we get
$$\begin{array}{c|ccccc|c}
\textrm{\small Vertices} & x_1 & x_2 & x_3 & x_4 & x_5 & \textrm{\small Intervals}\\
\hline
y_1 & 1 & 1 & 1 & 0 & 0 & [1,3]\\
y_2 & 1 & 0 & 0 & 1 & 0 & [4,5]\\
y_3 & 0 & 0 & 0 & 1 & 0 & [5,5]\\
\hline
\textrm{\small Intervals} & [1,4] & [2,3] & [3,3] & [4,6] & [6,6] &
\end{array}$$

\begin{prop}
Algorithm \ref{alg:big1} provides an interval representation of an interval bigraph $B$.
\end{prop} 

\vspace{-1.5em}\begin{pf*}{Proof.}
Let $B$ be an interval bigraph with biadjacency matrix $A$ is in R-C partition form. Let us denote the vertex corresponding to the $i^\textrm{th}$ row [$j^\textrm{th}$ column] of $\tilde{A}$ by $u_i$ [respectively, $v_j$]. Now suppose $u_iv_j=1$\footnote{For convenience, an entry of a matrix corresponding to, say, the vertex $u_i$ in the row and the vertex $v_j$ in the column will be denoted by, simply, $u_iv_j$.}. If $i\leqslant j$, then by Algorithm \ref{alg:big1}, $c_j=j\leqslant b_i$ and so $a_i=i\leqslant j\leqslant b_i$. Thus $[a_i,b_i]\cap [c_j,d_j]$ contains $j$ and hence it is nonempty. Again if $i>j$, then by Algorithm \ref{alg:big1}, $a_i=i\leqslant d_j$. So $c_j=j<i=a_i\leqslant d_j$ which implies $[a_i,b_i]\cap [c_j,d_j]\neq\emptyset$ as it contains $i$. Next let $u_iv_j=R$. Since $\tilde{A}$ is diagonalized, $i<j$. But then by Algorithm \ref{alg:big1}, $b_i<j$ and so $a_i\leqslant b_i<j=c_j\leqslant d_j$, i.e., $[a_i,b_i]\cap [c_j,d_j]=\emptyset$. Similarly, if $u_iv_j=C$, then $i>j$ and by Algorithm \ref{alg:big1}, it follows that $c_j=j\leqslant d_j<i=a_i\leqslant b_i$, i.e., $[a_i,b_i]\cap [c_j,d_j]=\emptyset$. Therefore Algorithm \ref{alg:big1} provides an interval representation of $\tilde{B}$ and hence of $B$, as $B$ is an induced subgraph of $\tilde{B}$. \hfill $\qed$
\end{pf*}

\section{Probe interval graphs}

Let $G=(V,E)$ be an undirected graph with an independent set $N\subseteq V$. Let $P=V\smallsetminus N$. We construct a bipartite graph $B=(U_1,U_2,E_1)$ with the partite sets $U_1=P$ and $U_2=V$ and two vertices $p\in U_1$ and $v\in U_2$ are adjacent in $B$ if and only if either $p=v$ (in $G$) or $pv\in E$ (i.e., $p$ and $v$ are adjacent in $G$). That is, $B$ is a bipartite graph whose biadjacency matrix is the submatrix $P\times V$ of the augmented adjacency matrix (cf.~page \pageref{'augmented'}) of $G$ consisting of all the columns, but only the rows corresponding to all the vertices of $P$. Henceforth we refer this graph as $B=(P,V,E_1)$.

We note that if $G=(V,E)$ is a probe interval graph with probes $P$ and nonprobes $N$, then the bipartite graph $B=(P,V,E_1)$ is necessarily an interval bigraph by the same assignment of intervals to the vertices as in $G$. But the following example shows that the above necessary condition is not sufficient.

\begin{exmp}\label{exmp:at}
{\em Consider the following graph, say, $G$. $G$ is not\footnote{Note that $G$ is a probe interval graph with probes $\set{a,c,d}$ and nonprobes $\set{b,e,f}$.} a probe interval graph with probes $P=\set{a,b,c,d}$ and nonprobes $N=\set{e,f}$ as neither $G$ nor the graph $G+ef$ (the graph obtained by joining the edge $ef$ to $G$) is an interval graph.} 

\vspace{1em}
\begin{figure}[h]
\begin{center}
\includegraphics*[scale=0.45]{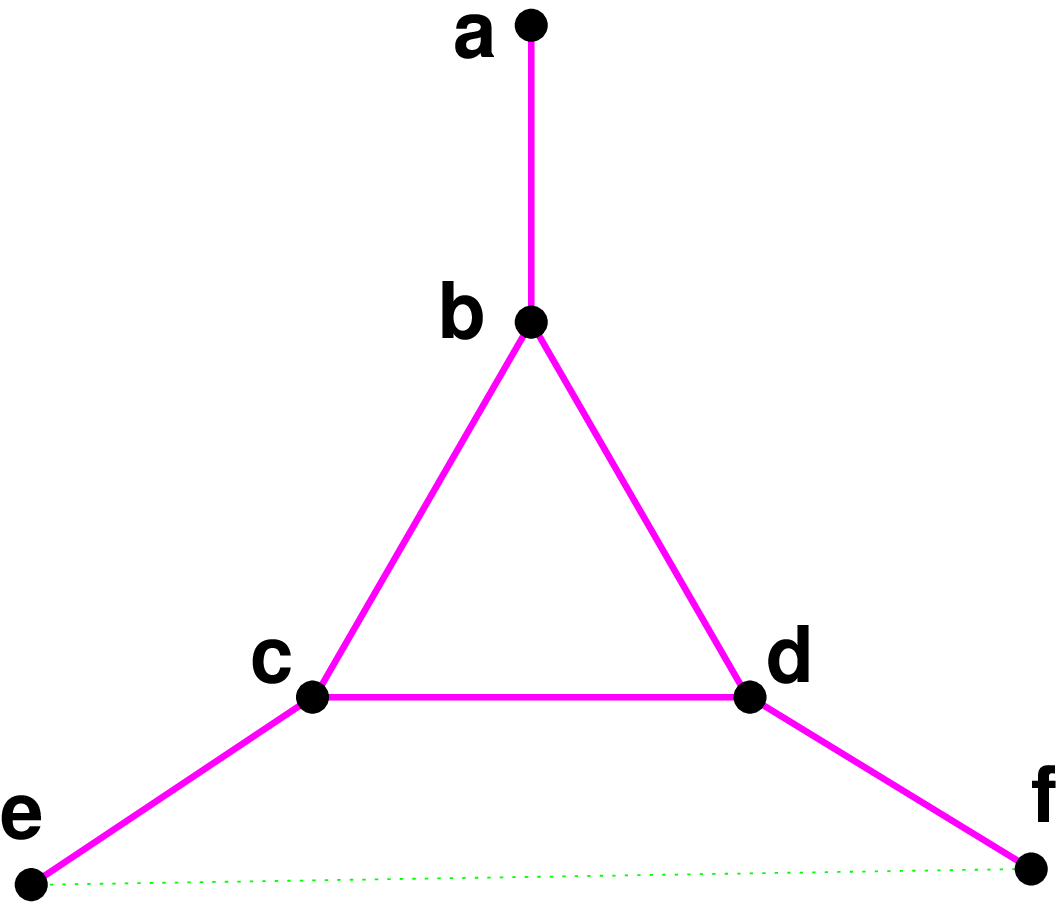}
\end{center}
\end{figure}

\vspace{1em}{\em But the biadjacency matrix of the bipartite graph $B=(P,V,E_1)$ has an R-C partition showing that $B$ is an interval bigraph.}
$$\begin{array}{c|cccccc}
\textrm{{\small vertices}} & a & b &  c & d & e & f \\
\hline
a & 1 & 1 & R & R & R & R \\
b & 1 & 1 & 1 & 1 & R & R \\
c & C & 1 & 1 & 1 & 1 & R \\
d & C & 1 & 1 & 1 & C & 1 
\end{array}$$
\end{exmp}

\begin{thm}\label{t:char1}
An undirected graph $G=(V,E)$ with an independent set $N\subseteq V$ is a probe interval graph with probes $P=V\smallsetminus N$ and nonprobes $N$ if and only if 
\begin{enumerate}
\item[{\bf (1)}] the bigraph $B=(P,V,E_1)$ is an interval bigraph and
\item[{\bf (2)}] there exists an R-C partition of the biadjacency matrix of $B$ which does not contain the following submatrix for any $p,q\in P$ and $n\in N$:\\[-1em]
\begin{equation}\label{eq:cfg}
\begin{array}{cc|cccccc}
\multicolumn{3}{c}{} & p && q && n\\ \cline{3-8}
p &&& 1 && 1 && R\\
q &&& 1 && 1 && C
\end{array}
\end{equation}
\end{enumerate}
\end{thm}

\begin{pf*}{Proof.}
Let $G=(V,E)$ be a probe interval graph with probes $P$ and nonprobes $N$. Then, as we observed earlier, the bipartite graph $B=(P,V,E_1)$ is an interval bigraph with the same assignment of intervals to all the vertices as in $G$.\footnote{Note that in the interval bigraph $B$, the same interval is assigned to every probe vertex $p$, both as a member of $P$ and as a member of $V$.} So its biadjacency matrix has an R-C partition by arranging vertices in the non-decreasing order of left end points of the intervals corresponding to them. Suppose for some $p,q\in P$ and some $n\in N$, there is a submatrix of the form (\ref{eq:cfg}). Let the intervals corresponding to $p,q$ and $n$ be $[a,b],[c,d]$ and $[l,r]$ respectively. Since $pn=R$, we have $a\leqslant b<l\leqslant r$ and since $qn=C$, we get that $l\leqslant r<c\leqslant d$ which imply $a\leqslant b<c\leqslant d$. Then it follows that the intervals $[a,b]$ and $[c,d]$ are disjoint. But this contradicts the fact that $pq=qp=1$. Thus we have the condition is necessary.

Conversely, let $G=(V,E)$ be an undirected graph with an independent set $N$ and $P=V\smallsetminus N$ such that $B=(P,V,E_1)$ is an interval bigraph and its biadjacency matrix, say, $M$ has an R-C partition which does not contain any submatrix of the form (\ref{eq:cfg}). We first show that it is possible to rearrange the columns of $M$ in such a way that the sequence of vertices of $P$ in the columns is same as that of in the rows of it and still the new matrix will have an R-C partition that does not contain a submatrix of the form (\ref{eq:cfg}). 

Suppose in $M$, $p,q\in P$ appear in the following manner:
$$\begin{array}{cc|cccc}
\multicolumn{3}{c}{} & q && p \\ \cline{3-6}
p &&&  && 1 \\
q &&& 1 &&  
\end{array}$$
Since $M$ is in R-C partition form, $pq$ cannot be $R$ or $C$. So we must have
$$\begin{array}{cc|cccc}
\multicolumn{3}{c}{} & q && p \\ \cline{3-6}
p &&& 1 && 1 \\
q &&& 1 && 1 
\end{array}$$
Now if the column of $p$ does not contain $R$, then all columns of $M$ left to that of $p$ also cannot contain $R$. Thus the column of $q$ can be placed just right to that of $p$ and the new matrix thus formed remains to be in the R-C partition form. Also since we did not change any $R,C$ or $1$ of the matrix $M$, the new matrix also does not contain a submatrix of the form (\ref{eq:cfg}). Again if all rows of $M$ for which the column of $p$ contains $R$ also have $R$ in the column of $q$, then we have $R$ in all the columns in between them. Thus, in this case also, shifting the column of $q$ just right to that of $p$ will not disturb the R-C partition of the matrix and will not invite any submatrix of the form (\ref{eq:cfg}). 

So suppose there exists $r\in P$ such that $rp=R$ and $rq=1$. Now since $rp=R$ and $rr=1$, the column of $r$ appears left to that of $p$ in $M$. Also since $pr=0$ (as $rp=0$) and $pp=1$, we have $pr=C$. But then we have
$$\begin{array}{cc|cccc}
\multicolumn{3}{c}{} & r && p \\ \cline{3-6}
r &&& 1 && R \\
p &&& C && 1 \\
q &&& 1 && 1
\end{array}$$
as $qr=rq=1$. Since this configuration is not allowed in an R-C partition, this case is not possible. Thus we can have the biadjacency matrix of $B$ in our desired form. Let us denote the matrix of this form by $M_1$ and the probe vertex corresponding to the $i^\textrm{th}$ row of $M_1$ by $p_i$.

Now let in the interval bigraph $B=(P,V,E_1)$ the interval corresponding to each $p_i\in P$ be $[a_i^\prime,b_i^\prime]$ and that corresponding to each $p_i\in V$ be $[a_i^{\prime\prime},b_i^{\prime\prime}]$. Further we assume that the interval representation of $B$ is obtained from $M_1$ by Algorithm \ref{alg:big1}. By Theorem 1 in \cite{SSW}, the assignment of the interval $[a_i^\prime + a_i^{\prime\prime},b_i^\prime + b_i^{\prime\prime}]\ =\ [a_i,b_i]$ (say) with each $p_i$ both as a member of $P$ as well as of $V$ yields an interval representation for the submatrix $P\times P$.

We replace the interval assignment of each $n_j\in N$ by $[l_j,r_j]$, where
\begin{equation}\label{eq:rj}
r_j=\left\{%
\begin{array}{l}
\min \Set{a_i}{p_in_j=C} - 1\\
\infty,\ \ \textrm{ if there is no $C$ in the column of $n_j$\footnotemark .} 
\end{array}\right .
\end{equation}
\footnotetext{Here the symbol $\infty$ stands for a sufficiently large positive integer which is greater than all the right end points assigned here.}
\begin{equation}\label{eq:lj}
l_j=\left\{%
\begin{array}{l}
\max \Set{b_i}{p_in_j=R} + 1\\
0,\ \ \textrm{ if there is no $R$ in the column of $n_j$.}
\end{array}\right .
\end{equation}

Let $n_k\in N$. First we ensure that $l_k\leqslant r_k$ indeed. 

If the column of $n_k$ does not contain $R$ [$C$], then $l_k=0$ [respectively, $r_k=\infty$] and in either case $l_k\leqslant r_k$. 

Next suppose the column of $n_k$ contains both $R$ and $C$. Let $p_in_k=R$ and $p_jn_k=C$. Since any $0$ below a $C$ in an R-C partition is $C$, we have $i<j$. Also due to our hypothesis we must have the following configuration:
$$\begin{array}{cc|cccccc}
\multicolumn{3}{c}{} & p_i && p_j && n_k\\ \cline{3-8}
p_i &&& 1 && 0 && R\\
p_j &&& 0 && 1 && C
\end{array}$$
Further since $p_jp_j=1$, we have $p_ip_j=R$ and $p_jp_i=C$. But then $b_i^\prime < a_j^{\prime\prime}$ and $b_i^{\prime\prime} < a_j^\prime$ and so $b_i=b_i^\prime + b_i^{\prime\prime} < a_j^\prime + a_j^{\prime\prime} = a_j$, which is true for any $i,j$ for which $p_in_k=R$ and $p_jn_k=C$. Thus 
$$\max \Set{b_i}{p_in_k=R}\ <\ \min \Set{a_j}{p_jn_k=C}.$$
Then by (\ref{eq:rj}) and (\ref{eq:lj}), we have $l_k<r_k$, as required.

Now we show that the new interval assignments agree with the given matrix $M_1$, i.e., for any $p_i\in P$ and $n_j\in N$, if $p_in_j=0$, then the intervals $[a_i,b_i]$ and $[l_j,r_j]$ do not intersect and if $p_in_j=1$, then the intervals $[a_i,b_i]$ and $[l_j,r_j]$ must intersect. That $[l_j,r_j]$ is disjoint from $[a_i,b_i]$ when $p_in_j=0$ (i.e., $R$ or $C$) is clear from the construction of (\ref{eq:rj}) and (\ref{eq:lj}).

Next suppose $p_kn_j=1$ for some $p_k\in P$ and $n_j\in N$. We show that $l_j\leqslant b_k$ and $a_k\leqslant r_j$. If there is no $R$ in the column of $n_j$, then $l_j=0<b_k$. Suppose $p_in_j=R$ for some probe $p_i\in P$. Since $p_kn_j=1$, $b_i^\prime < b_k^\prime$. Also if there is no $C$ in the  column of $p_k$, then $b_k^{\prime\prime}$ is greater than or equal to all right end points of vertices in the column of the matrix $M_1$ and hence $b_i^{\prime\prime}\leqslant b_k^{\prime\prime}$. Let there be a $C$ in the column of $p_k$, then it is below $p_kp_k$ (which is $1$). Suppose $p_tp_k=C$ for some $t>k$. Then $p_kp_t=R$ as $p_tp_t=1$ and $t>k$. So the column of $p_t$ appears right to that of $n_j$ as $p_kn_j=1$. But then $p_ip_t=R$ as $p_in_j=R$. Also since $p_ip_i=1$, the column of $p_i$ appears left to that of $n_j$ and hence also left to the column of $p_t$. Then $p_tp_i=C$ as $p_tp_t=1$. So we have 
\begin{equation}\label{eq:kit}
p_tp_i=C\ \textrm{ whenever }\ p_tp_k=C.
\end{equation}
Now if $i<k$, then it follows from (\ref{eq:kit}) that $b_i^{\prime\prime}\leqslant b_k^{\prime\prime}$. Let $i>k$, then $p_kp_i=1$ as $p_kn_j=1$ and $p_ip_i=1$. So $p_ip_k=1$. Then (\ref{eq:kit}) implies again $b_i^{\prime\prime}\leqslant b_k^{\prime\prime}$. Therefore $b_i=b_i^\prime + b_i^{\prime\prime} < b_k^\prime + b_k^{\prime\prime} =b_k$. Hence $\max \Set{b_i}{p_in_j=R} < b_k$ and so $l_j\leqslant b_k$, as required.

\begin{center}
$\begin{array}{cc|cccccccc}
\multicolumn{3}{c}{} & p_i && p_k && n_j && p_t\\ \cline{3-10}
p_i &&& 1 &&  && R && R\\
p_k &&&  && 1 && 1 && R\\
p_t &&& C && C && && 1\\
\end{array}$ \hspace{1in}
$\begin{array}{cc|cccccccc}
\multicolumn{3}{c}{} & p_k && p_i && n_j && p_t\\ \cline{3-10}
p_k &&& 1 && 1 && 1 && R\\
p_i &&& 1 && 1 && R && R\\
p_t &&& C && C && && 1\\
\end{array}$
\end{center}

Again if there is no $C$ in the column of $n_j$, then $r_j=\infty >a_k$. Suppose $p_in_j=C$ for some $p_i\in P$. Then $i>k$ as $p_kn_j=1$.  Then $a_k^\prime < a_i^\prime$. Also since probe vertices appear in the same sequence in the columns of $M_1$ as in the rows of it, we have $a_k^{\prime\prime} \leqslant a_i^{\prime\prime}$. Thus $a_k=a_k^\prime +a_k^{\prime\prime} < a_i^\prime +a_i^{\prime\prime}=a_i$. This implies $a_k<\min\Set{a_i}{p_in_j=C}$ and hence $a_k\leqslant r_j$. \hfill $\qed$
\end{pf*}

Now we proceed for another characterization of the adjacency matrix of a probe interval graph. Let $B=(X,Y,E)$ be a bipartite graph. For each $x\in X$, let $n(x)=\Set{y\in Y}{xy\in E}$ be the set of neighbors of $x$ . A {\em Ferrers bigraph} \cite{R} is a bipartite graph $B=(X,Y,E)$ in which sets of neighbors of vertices of $X$ are linearly ordered by set inclusion,\footnote{Similar condition for vertices of $Y$ is equivalent to this one, i.e., from this it follows that sets of neighbors of vertices of $Y$ are also linearly ordered by set inclusion.} i.e., there is a linear ordering of the vertices of $X=\set{x_1,x_2,\ldots ,x_n}$ (say) such that $n(x_i)\subseteq n(x_j)$ for all $i\leqslant j$. Another equivalent condition \cite{R} on a bipartite graph $B$ to be a Ferrers bigraph is that the biadjacency matrix of $B$ does not contain any $2\times2$ permutation matrix: 
$$\left(%
\begin{array}{cc}
1 & 0\\[-0.25em]
0 & 1
\end{array}
\right) \hspace{0.5in}
\textrm{ or } \hspace{0.5in}\left(%
\begin{array}{cc}
0 & 1\\[-0.25em]
1 & 0
\end{array}
\right).$$
It is well known that every bipartite graph is an intersection of a finite number of Ferrers bigraphs and the minimum such number is called its {\em Ferrers dimension}. The bipartite graphs of Ferrers dimension at most 2 were characterized by Cogis~\cite{C}. He called every $2\times2$ permutation matrix in a binary matrix a {\em couple} and defined an undirected graph $H(B)$, the graph {\em associated to a bipartite graph} $B$ as follows. The vertices of $H(B)$ correspond to the positions with entry $0$ in the biadjacency matrix, say, $A$ of $B$ and two such vertices are adjacent in $H(B)$ if and only if the corresponding $0$'s form a couple in the matrix $A$. Cogis proved that a bipartite graph $B$ is of Ferrers dimension at most 2 if and only if $H(B)$ is bipartite. In particular, a bipartite graph is an interval bigraph if and only if it is the intersection of two Ferrers bigraphs whose union is complete \cite{SDRW}. Moreover, when $B$ is an interval bigraph, any R-C partition of its biadjacency matrix provides a proper $2$-coloring (by colors $R$ and $C$) of vertices of $H(B)$. Thus it is important to note that, in this case, no two $R$'s [$C$'s] are in the same couple in the biadjacency matrix of $B$. 

Let $G=(V,E)$ be an undirected graph having $N\ (\subseteq V)$ as an independent set of vertices. Let $B_1$ be the bipartite graph whose biadjacency matrix is the augmented adjacency matrix (cf.~page \pageref{'augmented'}) of $G$. Now from the graph $H(B_1)$ delete the vertices corresponding to $0$'s in the submatrix $N\times N$ of the biadjacency matrix of $B_1$. Call it $H_1(B_1)$, the {\em reduced associated graph} of $B_1$.

\begin{thm}\label{t:char2}
Let $G=(V,E)$ be an undirected graph with an independent set $N\subseteq V$ and $B_1$ be the bipartite graph whose biadjacency matrix is the augmented adjacency matrix of $G$. Then $G$ is a probe interval graph with probes $P=V\smallsetminus N$ and nonprobes $N$ if and only if 
\begin{enumerate}
\item[{\bf (1)}] the bigraph $B=(P,V,E_1)$ is an interval bigraph and
\item[{\bf (2)}] the graph $H_1(B_1)$ is a bipartite graph and there is a bipartation of $H_1(B_1)$ that yields an R-C partition of $B$. 
\end{enumerate}
\end{thm}

\begin{pf*}{Proof.}
Let $G$ be a probe interval graph with probes $P$ and nonprobes $N$. Then by Theorem \ref{t:char1}, $B=(P,V,E_1)$ is an interval bigraph and there exists an R-C partition of $B$ which does not contain any submatrix of the form (\ref{eq:cfg}). We note that the graph $H_1(B_1)$ contains the graph $H(B)$ and only vertices of $H_1(B_1)$ which are not in $H(B)$ are the zeros of the biadjacency matrix $B_1$ at the positions $np$ for some $n\in N$ and $p\in P$ (i.e., the zeros of the submatrix $N\times P$). Now since $B$ is an interval bigraph, $H(B)$ is bipartite. Moreover the above R-C partition provides a proper $2$-coloring of vertices of $H(B)$ (by colors $R$ and $C$). Let us extend this coloring of vertices $H(B)$ to the vertices of $H_1(B_1)$ as follows:\\[-1em]
\begin{equation}\label{eq:np}
np=\left\{%
\begin{array}{l}
R, \qquad \textrm{ if } pn=C\\
C, \qquad \textrm{ if } pn=R.
\end{array}\right .
\end{equation}
Now if this assignment of colors provides a $2$-coloring of the vertices of $H_1(B_1)$, then we have nothing to prove. If not, then there exist couples of the forms:
$$\left(%
\begin{array}{cc}
1 & R\\[-0.25em]
R & 1
\end{array}
\right) \hspace{0.5in}
\textrm{ or } \hspace{0.5in}\left(%
\begin{array}{cc}
1 & C\\[-0.25em]
C & 1
\end{array}
\right).$$
in the biadjacency matrix of $B_1$ where none of the zeros ($R$ or $C$) belongs to the submatrix $N\times N$. Also since vertices of $H(B)$ is properly $2$-colored (by $R$ or $C$), at least one of the two rows of these couples must corresponds to a nonprobe (i.e., these couples cannot lie fully in the submatrix $P\times V$). So the following three cases may arise for couples of the first type (containing $R$'s):
$$\begin{array}{cc|cccc}
\multicolumn{3}{c}{} & p && q \\ \cline{3-6}
m &&& 1 && R \\
n &&& R && 1 
\end{array}\hspace{0.5in} 
\begin{array}{cc|cccc}
\multicolumn{3}{c}{} & p && q \\ \cline{3-6}
r &&& 1 && R \\
n &&& R && 1 
\end{array}\hspace{0.5in}
\begin{array}{cc|cccc}
\multicolumn{3}{c}{} & p && n \\ \cline{3-6}
q &&& 1 && R \\
n &&& R && 1 
\end{array}$$
where $p,q,r\in P$ and $m,n\in N$. The first case implies the existence of
$$\begin{array}{cc|cccc}
\multicolumn{3}{c}{} & m && n \\ \cline{3-6}
p &&& 1 && C \\
q &&& C && 1 
\end{array}$$
in the biadjacency matrix, say, $M$ of $B$ which is not possible. The second one again forces the following in $M$:
$$\begin{array}{cc|cccc}
\multicolumn{3}{c}{} & r && n \\ \cline{3-6}
p &&& 1 && C \\
q &&& X && 1 
\end{array}$$
where $X=R$ or $C$. Clearly $X\neq C$. Suppose $X=R$. But then we have $qr=R=rq$ and consequently the couple 
$$\begin{array}{cc|cccc}
\multicolumn{3}{c}{} & q && r \\ \cline{3-6}
q &&& 1 && R \\
r &&& R && 1 
\end{array}$$
in $M$, which is a contradiction. So finally we consider the last one. In this case we get the submatrix
$$\begin{array}{cc|cccccc}
\multicolumn{3}{c}{} & q && p && n \\ \cline{3-8}
q &&& 1 && 1 && R \\
p &&& 1 && 1 && C 
\end{array}$$
in $M$ which is of the form (\ref{eq:cfg}) and so is forbidden as we mentioned at the beginning of the proof. The proof for the couples of other type (contaning $C$'s) is similar and hence omitted. Therefore $H_1(B_1)$ is bipartite and there is a bipartation of it which yields an R-C partition of $B$.

Conversely, let the conditions (1) and (2) be satisfied. So we have the bigraph $B=(P,V,E_1)$ is an interval graph, the graph $H_1(B_1)$ is bipartite and there is a bipartation of $H_1(B_1)$ which gives an R-C partition of $B$. We show that such an R-C partition of $B$ cannot contain any submatrix of the form (\ref{eq:cfg}). Then it will follow that $G$ is a probe interval graph by Theorem \ref{t:char1}. 

Now if the R-C partition of $B$ has a submatrix of the form (\ref{eq:cfg}), then we have the following submatrix:
$$\begin{array}{cc|cccccccc}
\multicolumn{3}{c}{} & p &&& q &&& n \\ \cline{3-10}
p &&& 1 &&& 1 &&& R \\
q &&& 1 &&& 1 &&& C \\
n &&& X &&& Y &&& 1
\end{array}$$
in the biadjacency matrix of $B_1$, where $X,Y\in\set{R,C}$.\footnote{Denoting all the vertices of one partite set of $H_1(B_1)$ by $R$ and those of the other by $C$ such that this yields the R-C partition of $B$.} But $X$ cannot be either $R$ or $C$ as we have the following couples in the above submatrix:
$$\begin{array}{cc|cccc}
\multicolumn{3}{c}{} & p && n \\ \cline{3-6}
p &&& 1 && R \\
n &&& X && 1 
\end{array}\qquad \textrm{ and }\qquad \begin{array}{cc|cccc}
\multicolumn{3}{c}{} & q && n \\ \cline{3-6}
q &&& 1 && C \\
n &&& X && 1. 
\end{array}$$
This contradiction proves that the above R-C partition cannot contain any submatrix of the form (\ref{eq:cfg}), as required.\hfill $\qed$
\end{pf*}

Finally, in the following we show that if we add a loop at every probe vertex of a probe interval graph, then the Ferrers dimension of the corresponding symmetric bipartite graph is at most $3$. We know that an interval bigraph is of Ferrers dimension at most $2$, but the converse is not true. Below we show that the property of being of Ferrers dimension at most $2$ is also a sufficient criterion for a graph to be an interval graph.

\begin{prop}
An undirected graph $G=(V,E)$ is an interval graph if and only if the Ferrers dimension of the corresponding bipartite graph $B$, whose biadjacency matrix is the augmented adjacency matrix of $G$, is at most $2$.
\end{prop}

\begin{pf*}{Proof.}
From the quasi-linear ones property of the augmented adjacency matrix of an interval graph it is clear that the $0$'s in the upper triangle and those in the lower triangle form two Ferrers digraphs whose union is $\overline{G}$, the complement of $G$. This proves the direct part.

Conversely, Cogis \cite{C} proved that a bipartite graph $B_1$ is of Ferrers dimension at most $2$ if and only if its associated graph $H(B_1)$ is bipartite. In fact he proved that if $H(B_1)$ has nontrivial components $H_1, H_2,\ldots ,H_k$ and has $I=\set{I_1,I_2,\ldots ,I_m}$ as its isolated vertices, then there is a $2$-coloring $(R_i,C_i)$ of $H_i\ (i=1,2,\ldots ,k)$ so that $R=R_1\cup R_2\cup\cdots\cup R_k\cup I$ and $C=C_1\cup C_2\cup\cdots\cup C_k\cup I$ are two Ferrers bigraphs whose union is $\overline{G}$. Clearly, if there is no isolated vertex, i.e., $I=\emptyset$, then $\overline{G}$ is decomposed into disjoint Ferrers bigraphs. 

Let $F_1$ and $F_2$ be two Ferrers bigraphs whose union is $\overline{G}$, i.e., $\overline{G}=F_1\cup F_2$. Let $A$ be the augmented adjacency matrix of $G$ and so the biadjacency matrix of $B$. Let $uv=0$. Then $vu=0$ and the couple
$$\begin{array}{cc|cccc}
\multicolumn{3}{c}{} & u && v\\ \cline{3-6}
u &&& 1 && 0 \\
v &&& 0 && 1 \\
\end{array}$$
in the matrix $A$ shows that the two $0$'s at positions $uv$ and $vu$ are adjacent in $H(B)$. Let $uv\in F_1$ so that $vu\in F_2$. Thus every $0$ in the matrix $A$ belongs to a non-trivial component of $H(B)$, which implies that $H(B)$ has no isolated vertex. Hence, as noted earlier, the two Ferrers bigraphs $F_1$ and $F_2$ are disjoint. So $B$ is an interval bigraph and consequently by Theorem 1 of \cite{SSW} we have $G$ is an interval graph.\hfill $\qed$
\end{pf*}

Let $G$ be a probe interval graph. Let us add a loop at every probe vertex of $G$ and denote the graph thus obtained by $\widehat{G}$.

\begin{cor}
Let $G=(V,E)$ be a probe interval graph. Then the Ferrers dimension of the bipartite graph, whose biadjacency matrix is the adjacency matrix of $\widehat{G}$, is at most $3$.
\end{cor}

\begin{pf*}{Proof.}
Let $G$ be a probe interval graph with probes $P$ and nonprobes $N$. Let $G_1=(V,E_1)$ be the interval graph with the same assignment of intervals to all the vertices as in $G$. Let $B$ be the bipartite graph whose biadjacency matrix is the adjacency matrix of $\widehat{G}$\footnote{Note that, in the adjacency matrix of $\widehat{G}$, $pp=1$ for each probe vertex $p$ of $G$.} and $B_1$ be the bipartite graph whose biadjacency matrix is the augmented adjacency matrix of $G_1$. Then by the above theorem, we have $B_1$ is of Ferrers dimension at most $2$ and so $B_1=F_1\cap F_2$ for some Ferrers bigraph $F_1$ and $F_2$ such that $F_1\cup F_2$ is complete. Also the bipartite graph, whose biadjacency matrix is the following matrix, is a Ferrers bigraph, say, $F_3$.
$$\begin{array}{cc|ccc|ccc|c}
\multicolumn{3}{c}{} & P & \multicolumn{2}{c}{} &N& \multicolumn{2}{c}{}\\ \cline{3-8}
P &&& \mathbf{1} &&& \mathbf{1} &&\\ \cline{3-8}
N &&& \mathbf{1} &&& \mathbf{0} &&\\ \cline{3-8}
\end{array}$$

Thus we have $B=F_1\cap F_2\cap F_3$, as required.\hfill $\qed$
\end{pf*}

\begin{ack}
The authors are grateful to the learned referees for their meticulous reading and valuable suggestions which have definitely improved the paper.
\end{ack}

\end{document}